\documentclass[sigconf]{acmart}

\AtBeginDocument{%
  \providecommand\BibTeX{{%
    \normalfont B\kern-0.5em{\scshape i\kern-0.25em b}\kern-0.8em\TeX}}}
\usepackage{pgfplots} 
\usepackage{hyperref}
\usepackage{enumitem}
\usepackage{stfloats}
\usepackage{xcolor}
\usepackage{url}
\usepackage{soul}
\usepackage{colortbl}
\usepackage{subfigure}
\usepackage{caption}
\usepackage{subcaption}
\usepackage{xcolor}
\usepackage{refcount}
\usepackage{float}
\usepackage{siunitx}
\usepackage{booktabs}
\usepackage{graphicx}
\usepackage{multirow}
\usepackage{makecell} 
\usetikzlibrary{tikzmark,calc}
\usepackage{calc}
\usepackage{booktabs}
\usepackage{microtype}
\usepackage{array}
\usepackage{graphicx}
\usepackage{tikz}
\usetikzlibrary{shapes,arrows.meta,positioning,calc,backgrounds}

\definecolor{vlightgray}{gray}{0.9}

\hypersetup{
 colorlinks=true,   
 citecolor=black,
 urlcolor=blue,
 linkcolor=black,
 bookmarksnumbered=true,     
 bookmarksopen=true,         
 bookmarksopenlevel=1,          
 pdfstartview=Fit,           
 pdfpagemode=UseOutlines,    
 pdfpagelayout=TwoPageRight 
}

\newcolumntype{P}[1]{>{\raggedright\arraybackslash}p{#1}}

\newcommand{\pQuote}[1]{{\color{TekheletPurple}\textit{#1}}}

\newif\ifdraft
\draftfalse 

\newif\ifrevising
   \revisingfalse 

 \newcommand{\deleted}[1]{{\ifrevising{\relax}\else\relax\fi}}
\usepackage{framed}
\usepackage{xcolor}

\definecolor{TekheletPurple}{HTML}{572F96}

\newenvironment{takeawayBox}{%
  \MakeFramed{\advance\hsize-\width\FrameRestore}%
}{\endMakeFramed}

\newenvironment{implicationBox}{%
  \MakeFramed{\advance\hsize-\width\FrameRestore}%
}{\endMakeFramed}

\makeatletter
\providecommand\nopunct{\@addpunct{}}
\makeatother

\title{What's Beyond Copilot?\\
22 AI Systems Developers Want Built}

\title{Looking Beyond Copilot:\\
22 AI Systems Developers Want Built}

\title{To Copilot and Beyond:\\
22 AI Systems Developers Want Built}

\makeatletter
\renewcommand{\@authorsaddresses}{}
\makeatother

\author{%
  Rudrajit Choudhuri$^{1}$ \quad
  Christian Bird$^{2}$ \quad
  Carmen Badea$^{2}$ \quad
  Anita Sarma$^{1}$
  \country{}
}
\affiliation{%
  $^{1}$Oregon State University, OR, USA. Email: \{choudhru, anita.sarma\}@oregonstate.edu \\
  $^{2}$Microsoft Research, WA, USA. Email: \{cbird, cabadea\}@microsoft.com%
  \country{}
}

\setcopyright{none}
\acmConference[]{Oregon State University \& Microsoft Research}{USA}{2026}

\renewcommand{\shortauthors}{Choudhuri et al.}

\begin{abstract}

Developers spend roughly one-tenth of their workday writing code, yet most AI tooling targets that fraction. This paper asks what should be built for the rest. We surveyed 860 Microsoft developers to understand where they want AI support, and where they want it to stay out. Using a human-in-the-loop, multi-model council-based thematic analysis, we identify 22 AI systems that developers want built across five task categories. For each, we describe the problem it solves, what makes it hard to build, and the constraints developers place on its behavior.

Our findings point to a growing \textit{right-shift burden} in AI-assisted development: developers wanted systems that embed quality signals earlier in their workflow to keep pace with accelerating code generation, while enforcing explicit authority scoping, provenance, uncertainty signaling, and least-privilege access throughout. This tension reveals a pattern we call \textit{``bounded delegation''}: developers wanted AI to absorb the assembly work surrounding their craft, never the craft itself. That boundary tracks where they locate professional identity, suggesting that the value of AI tooling may lie as much in where and how precisely it stops as in what it does.

\end{abstract}

\begin{document}

\maketitle

\section{Introduction}

When asked what software developers do, most people picture someone writing code. AI tooling ventures have largely followed the same intuition, concentrating investment on code generation~\cite{chen2021evaluating, ziegler2024measuring}. But studies of how developers actually spend their time tell a different story: writing code accounts for roughly one-tenth of their day~\cite{kumar2025time, meyer2019today}. The rest, spent debugging production incidents, navigating compliance, onboarding teammates, keeping documentation from drifting, reviewing changes, and translating technical decisions for non-technical stakeholders, is where most of the work lives~\cite{kumar2025time, obi2025identifying}, and where AI support remains comparatively sparse.

This mismatch is creating a compounding problem in software development: AI-assisted code generation is accelerating the part of the lifecycle that was already comparatively fast, while expanding the parts that are slow, messy, and human-intensive (e.g., review, verification, testing, incident triage)~\cite{baltes2026endless}. Reviewers face more code with less provenance~\cite{choudhuri2025needs}. On-call engineers debug systems whose logic was never fully understood by the team that shipped them~\cite{pearce2022asleep}. Documentation falls further behind because the code it describes changes faster than anyone can track~\cite{yamasaki2026writes}.
Emerging evidence suggests this dynamic is already producing measurable strain. AI-generated ``workslop,'' output that appears useful but lacks substance, forces recipients to interpret, correct, or redo the work~\cite{niederhoffer2025ai}, undermining the very productivity gains that AI was supposed to deliver. The downstream costs compound: developer fatigue and burnout~\cite{miller2025maybe, feng2025gains}, and a growing accumulation of AI-induced technical debt~\cite{liu2026debt, choudhuri2025needs}.

Designing tools that address this problem requires knowing what developers actually need, and that requires their voice. Decades of work-design research show that automation-first approaches cap gains at labor arbitrage; augmentation is what expands them~\cite{brynjolfsson2022turing, allan2019outcomes, bailey2019review, autor2022labor}. When workers' perspectives inform tool design, organizations achieve more sustainable improvements in both productivity and well-being~\cite{hackman1976motivation, trist1951some}.

Yet most studies of AI in software engineering examine current adoption patterns, productivity effects, or task-level exposure~\cite{russo2024navigating, afroz2025developer, khemka2024toward, butler2025dear, feng2025gains, choudhuri2025guides, lambiase2025exploring, kumar2025time}. 
They tell us where developers use AI and why they adopt or resist it. They do not tell us what should be built next, what those systems should do, or what they should never do on their own. This paper provides that study. We ask:

\

\textbf{RQ}: \textit{What AI systems do developers want built, and what conditions do they place on those systems for them to be acceptable?}

\

The data comes from a large-scale survey of 860 Microsoft developers reported in~\cite{choudhuri2025ai}, which investigated how developers' cognitive appraisals of their daily work shape their openness to and use of AI. This paper asks the complementary question: what, concretely, should be built? Using a human-in-the-loop, multi-model council-based thematic analysis~\cite{wu2026council, tai2024examination} of the survey's open-ended responses, we identified 22 AI systems across five software development task categories that developers want built. For each, we describe the problem developers want solved, what makes it hard to build, and the constraints developers place on its behavior.

Developers' needs fell overwhelmingly on the verification side of the workflow: they wanted systems to embed quality signals earlier, at authorship time, at the point of change, to keep pace with accelerating code generation. Across these systems, developers enforced four guardrails: (1) explicit authority scoping, (2) provenance, (3) uncertainty signaling, and (4) least-privilege access.

Our findings reveal an underlying pattern we call \textit{``bounded delegation''}: developers wanted AI to \textit{absorb the assembly work surrounding their craft, never the craft itself}. That boundary reflected where they located professional identity, accountability, and ownership in their work. It did not reflect capability; developers drew it even for tasks they acknowledged AI could plausibly handle, suggesting it is unlikely to move simply because models improve. This paper helps chart where that boundary lies and why developers defend it, and leaves with the question: \textit{what, then, should AI systems for software engineering be designed to protect?}

\section{Related Work}

As AI increasingly absorbs routine coding work, developers are left to navigate the broader software lifecycle (e.g., design, architecture, planning, compliance, and operations)---areas where AI support remains relatively sparse and poorly understood~\cite{bird2022taking, kumar2025time}. Understanding where support is most needed, and what it would take to build it, has therefore emerged as a focal research topic.

Prior studies have examined factors shaping AI adoption in software engineering (SE)~\cite{choudhuri2025guides, russo2024navigating, banh2025copiloting, miller2025maybe, butler2025dear}. Workflow compatibility has been shown to predict early adoption strongly: tools that fail to align with developers’ existing practices are often abandoned regardless of their capabilities~\cite{russo2024navigating}. Trust further shapes usage, influenced by both system characteristics and individual dispositions toward AI~\cite{wang2023investigating, johnson2023make, cheng2023would, choudhuri2025guides}. In particular, reliability, transparency, goal maintenance, and provenance are critical for developers’ trust, yet remain insufficiently supported in available tools~\cite{choudhuri2025needs, johnsonfacilitating}.

More recent work has shifted to task-level analyses of AI use~\cite{khemka2024toward, pereira2025exploring, kumar2025time}, revealing that developers use AI for implementation tasks while retaining judgment in architecture and review~\cite{bird2022taking, kalliamvakou_2024}. They also identify unmet demand concentrated in testing, debugging, documentation, and compliance~\cite{khemka2024toward}, and link toil-heavy tasks to reduced developer satisfaction and productivity, highlighting them as promising targets for AI support~\cite{kumar2025time}. Complementary findings indicate that developers are more receptive to AI for artifact manipulation and information retrieval~\cite{lambiase2025exploring}, and less so for collaborative or creative work~\cite{pereira2025exploring}.

The survey underlying this paper, reported in~\citet{choudhuri2025ai}, provides the first empirically validated mapping of developers' daily work experiences to their AI adoption patterns and priorities for Responsible AI support.
Using cognitive appraisal theory~\cite{lazarus1991emotion, roseman2001appraisal} across a large-scale survey of developers, the study showed that perceived task value, accountability, and demands increase both openness to and use of AI. At the same time, identity alignment produces a dual effect: lower openness but higher usage when AI complements developers’ sense of meaningful work. The work further introduced an AI openness vs. usage landscape that highlights an opportunity space for tooling investments.

What this work established was \textit{where} developers want AI support and \textit{why}. Yet, it left open a critical question: \textit{what should be built?} 
In this study, we address this gap through a parallel qualitative analysis of developers' free-text responses from the same survey. From this analysis, we derive a grounded set of AI systems developers want built, and the constraints they place on those systems for them to be acceptable in practice.

\section{Method}

To address our RQ, we surveyed software developers at Microsoft. With over 60K developers spanning domains, roles, processes, stakeholder contexts, and geographies, and with sustained exposure to both mature and emerging AI tools, the organization provided a rich and diverse setting for this study.

\subsection{Source Survey}

The data for this study comes from an IRB-approved survey of 860 Microsoft developers conducted in July 2025~\cite{choudhuri2025ai}. The survey captured how developers cognitively appraise their work and how these appraisals relate to their AI adoption patterns and Responsible AI (RAI) priorities across SE tasks. It also included two open-ended questions per task category (see Tab.~\ref{tab:categories}), asking where developers most wanted AI support and where they did not. The companion paper analyzes these responses to explain \textit{why} developers seek or limit AI, using cognitive appraisals as the explanatory framework.

This paper addresses a complementary question: \textit{what, concretely, should be built?} We conduct a parallel qualitative analysis of the open-ended responses to derive a grounded set of systems developers want built.

\subsubsection{Survey Design}

The survey used a grounded taxonomy of software engineering tasks (Table~\ref{tab:categories}), constructed by integrating work-week studies of developer activities~\cite{meyer2019today,kumar2025time} with large-scale surveys on AI adoption~\cite{khemka2024toward,choudhuri2025guides}, and refined through pilot sessions with developers and SE researchers outside the research team.

\begin{table}[bhtp]
\small
\caption{Grounded taxonomy of SE tasks~\cite{kumar2025time, meyer2019today, choudhuri2025guides, khemka2024toward}}
\label{tab:categories}
\centering
\begin{tabular}{>{\raggedright\arraybackslash}m{2.3cm} >{\raggedright\arraybackslash}m{5.7cm}}
\hline
\textbf{Category} & \textbf{Tasks} \\
\hline \hline
\rowcolor{TekheletPurple!20!white} Development & Coding/Programming, Bug Fixing/Debugging, Performance Optimization, Refactoring \& Maintenance/Updates, AI Integration\\ 
Design \& Planning & System Design, Requirements Engineering, Project Planning \& Management \\
\rowcolor{TekheletPurple!20!white} Quality \& Risk Management & Testing \& Quality Assurance; Code Review/Pull Requests; Security \& Compliance \\
Infrastructure \& Operations & DevOps(CI/CD); Environment Setup \& Maintenance; Infrastructure Monitoring; Customer Support \\
\rowcolor{TekheletPurple!20!white} Meta-work (Collaboration/Knowledge Building) & Documentation; Client/Stakeholder Communication; Mentoring \& Onboarding; Learning; Research \& Brainstorming \\
\hline
\end{tabular}
\end{table}

Participants selected 2--3 task categories that best reflected their current work and answered questions for those categories. To reduce fatigue, the meta-work category (applicable to all developers) was excluded from the initial selection and shown only if a participant selected two categories; thus, no participant completed more than three category blocks.

Within each selected task category, participants completed Likert-scale items on task appraisals (value, identity, accountability, demands), AI openness, and RAI priorities (reported in the companion study~\cite{choudhuri2025ai}), followed by two open-ended questions:

\begin{enumerate}
    \item \textit{Opportunity:} ``Where do you want AI to play the biggest role in [task-category] activities?''
    \item \textit{Constraint:} ``What aspects do you not want AI to handle in [task-category] activities and why?''
\end{enumerate}

These questions elicit desired capabilities and unmet needs (opportunity), as well as boundary conditions (constraints), and constitute the primary data for this paper. The unit of analysis is a single respondent’s answer to one question within a task category; responses could be assigned multiple codes.

\subsubsection{Data collection} The survey was distributed via email to 8,000 developers, sampled across product groups, roles, and geographies. One reminder was sent after one week. Participation was voluntary and anonymous.

We received 1,193 responses (response rate: 14.86\%), consistent with prior large-scale SE surveys~\cite{punter2003conducting,storey2019towards}. We excluded incomplete responses ($n = 152$), patterned responses (e.g., straight-lined or repetitively alternating; $n = 59$), attention-check failures ($n = 98$), and participants reporting no AI experience ($n = 24$), resulting in 860 valid responses spanning six continents and diverse SE and AI experience. Respondent demographics are detailed in~\cite{choudhuri2025ai}.

The 860 participants provided open-ended responses across five task categories: Development ($n = 816$), Design \& Planning ($n=548$), Meta-Work ($n = 532$), Quality \& Risk Management ($n = 401$), and Infrastructure \& Operations ($n = 283$), yielding 2,580 response sets in total. Not all participants who completed the Likert-scale items provided substantive open-ended responses; the response counts in Sec.~\ref{sec:systems} reflect those who did. 
Participants are referenced hereafter as P1--P860.

\subsection{Data Analysis}
\label{sec:analysis}

\begin{figure*}[!thb]
\small
\centering
\includegraphics[width=0.95\linewidth]{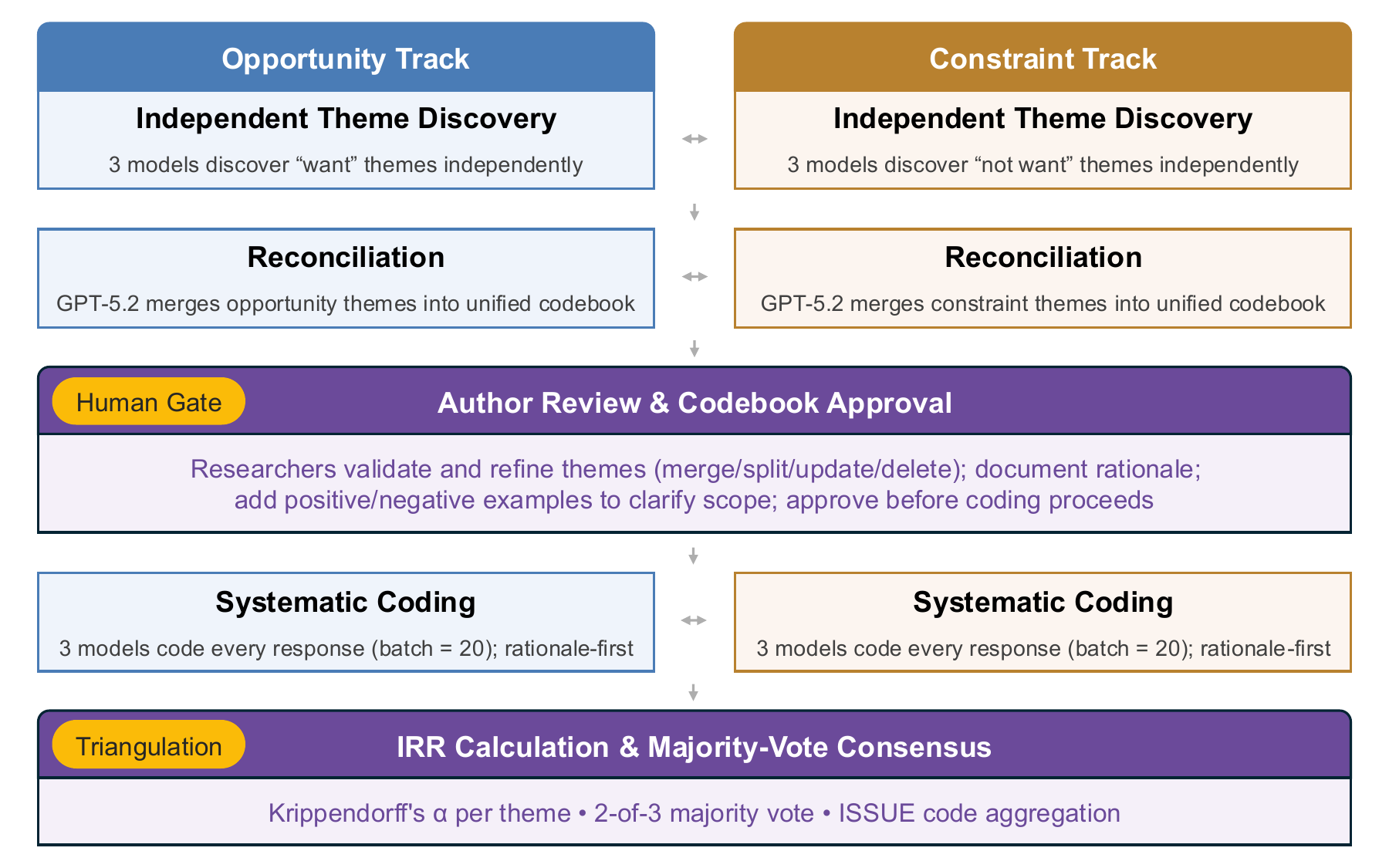}
\caption{\small Analysis pipeline overview. Three frontier models independently discover themes from developers' open-ended survey responses across two tracks. A reconciliation step produces a unified codebook per track. Researchers validate and refine themes before systematic coding proceeds. All three models then code each response against the approved codebook, providing a rationale before assigning codes. Inter-rater reliability is computed per theme using Krippendorff's $\alpha$\cite{gwet2014handbook}; final assignments use 2-of-3 majority vote.}
\label{fig:pipeline}
\end{figure*}

We analyzed developers' open-ended responses using a human-in-the-loop, multi-model council-based analysis pipeline (outlined in Fig.~\ref{fig:pipeline}), following reflexive thematic analysis~\cite{braun2006using,braun2022conceptual}. We designed this approach to balance scale with interpretive rigor in large-scale qualitative analysis~\cite{tai2024examination}. Both the opportunity (what developers wanted) and constraint (what they did not want) tracks followed identical processes with track-specific prompts and codebooks (provided in the supplemental package~\cite{supplemental}).

\subsubsection{\textbf{Stage 1: Independent theme discovery.}}
We used three frontier models for theme discovery: GPT-5.2 (OpenAI), Gemini 3.1 Pro (Google), and Claude Opus 4.6 (Anthropic).
We intentionally selected models from three distinct provider families to reduce model-specific blind spots, as different models could have exhibited distinct inductive biases and failure modes~\cite{wu2026council}. Agreement across model families therefore provides stronger convergent evidence than agreement within a single model family, where shared training data and optimization objectives can produce shared errors~\cite{bommasani2021opportunities, liang2022holistic}. 

Each model independently analyzed the full set of responses for a given category (e.g., Infra \& Ops) and discovered candidate themes with supporting participant IDs. No model had visibility into the others’ outputs. Prompts emphasized problem-focused, specific, and actionable themes grounded in the data (see~\cite{supplemental}). All models were configured with extended reasoning modes to reduce superficial pattern matching in open-ended text classification~\cite{wei2022chain}.


\subsubsection{\textbf{Stage 2: Codebook reconciliation.}}

We reconciled the independently discovered theme sets using GPT-5.2. The model was instructed to (1) merge overlapping themes representing the same underlying concept, (2) retain single-model themes only if supported by at least three distinct participant responses, and (3) remove themes not grounded in concrete developer needs or too sparsely evidenced to support reliable coding. We used a single model for this stage as it was a consolidation task rather than a discovery task.

Each resulting theme recorded its \texttt{source\_models} (which models proposed it) and \texttt{source\_codes} (original labels). This facilitated auditing of theme lineage and signaled convergent validity: themes independently identified by multiple models carried stronger evidentiary support; the provenance made that distinction visible at the human review gate.

\subsubsection{\textbf{Stage 3: Author review and codebook approval}}
\label{sec:human-gate}

Two researchers independently reviewed the reconciled codebooks before coding began. Each of them examined every proposed theme alongside all participant responses cited as supporting evidence. For each theme, they assessed: (1) fidelity: whether the label accurately captured what participants described; (2) grounding: whether cited responses genuinely supported the theme or reflected model over-extension; and (3) distinctness: whether boundaries between themes were sufficiently clear to support reliable coding. To further clarify theme boundaries, each researcher documented positive examples (clear instances of the theme) and negative examples (responses that appeared similar but fell outside its boundary).

The researchers then met to compare assessments using a negotiated agreement protocol~\cite{creswell2016qualitative}. Themes were split when too broad, merged when overlapping, added when concepts were missing, and removed when insufficiently grounded in the data. Disagreements were resolved until consensus was reached. Systematic coding proceeded only after final codebook approval.

\subsubsection{\textbf{Stage 4: Systematic coding.}}
\label{sec:coding}
All three models independently coded every response against the finalized codebook. Each coding instance required the model to write a chain-of-thought rationale before assigning codes~\cite{wei2022chain}, producing an auditable decision trail for human review.

To improve contextual accuracy, each response was paired with the participant’s answer to the complementary question (opportunity vs.\ constraint) within the same task category. This enabled the identification of misclassified or ambiguous responses and improved the interpretation of underspecified answers.
For example, in the meta-work category, one participant answered the question about where they wanted AI help with simply ``Everything.'' In isolation, this might have been coded as a desire for AI support across all meta-work tasks. However, their paired response to where they did \textit{not} want AI help was ``Communicating with coworkers,'' which revealed a clear exception. In another case, a participant said they wanted AI help with ``IcMs, development, refactoring and testing,'' but clarified in the paired response that ``Refactoring a system-wide feature, AI won't do it correctly.'' This additional context helped us code requests more precisely.

Models were restricted to the approved codebook; no novel codes were permitted. This constraint prevented code drift, which can occur when the same model processes large volumes of responses sequentially and begins inferring patterns not sanctioned by the codebook. Multi-coding was permitted to capture the full semantic breadth of each response.

Rather than discarding problematic responses, models flagged data quality issues using \texttt{ISSUE\_*} codes. 
Approximately 11\% responses exhibited such issues: misplaced answers (\texttt{ISSUE\_WRONG\_FIELD}), back-references unintelligible on their own (\texttt{ISSUE\_BACK\_REFERENCE}), or terse non-responses such as ``N/A'' (\texttt{ISSUE\_NON\_RESPONSE}). 
Models could introduce additional \texttt{ISSUE\_*} codes when the predefined set was insufficient.
Flagged responses were excluded from prevalence counts but retained for contextual interpretation of other responses from the same participant.

\vspace{-2mm}

\subsubsection{\textbf{Stage 5: Inter-rater reliability and consensus.}}
\label{sec:irr}

We assessed inter-rater reliability (IRR) among three models using Krippendorff’s $\alpha$~\cite{gwet2014handbook}, complemented by pairwise Cohen’s $\kappa$~\cite{gisev2013interrater} for each model pair (GPT--Gemini, GPT--Opus, Gemini--Opus) and three-rater percent agreement per theme. $\alpha$ captured overall agreement corrected for chance; $\kappa$ revealed whether alignment was uniform across pairs or concentrated in one; percent agreement identified full consensus. 
Across themes, IRR values ranged from 0.81 to 0.97 (mean = 0.94), indicating high reliability~\cite{gwet2014handbook}.
Themes below the threshold were flagged for review and subsequently refined, merged, or excluded through team consensus. Final theme assignments used majority vote: at least 2 of 3 models had to agree for a theme to be assigned to a response, applied independently per response and per theme. 

Responses were excluded from analysis if two or more models flagged any \texttt{ISSUE\_*} code, preventing over-filtering by any single model while still capturing systematic data-quality issues. Furthermore, two researchers spot-checked assignments against model rationales to ensure fidelity to the finalized codebook.

Finally, we synthesized related themes into system descriptions by consolidating, for each proposed system, the problems developers described, the capabilities they expected, and the constraints they imposed on its behavior, supported by representative participant quotes (see ~\cite{supplemental}).


\textbf{Reflexivity Note.} This pipeline is designed to augment, rather than replace, qualitative analysis. Models discover candidate patterns; researchers interpret and validate them. The resulting themes and system descriptions are grounded in verbatim responses that readers can inspect throughout Sec.~\ref{sec:systems}. We discuss the methodological implications of AI-assisted qualitative coding in Sec.~\ref{sec:research}.

\subsection{Limitations}

\textbf{Construct validity.} We report stated preferences rather than predicting adoption. Developers may describe capabilities they would not sustain in practice. We treat this as motivation and a direction for future work: claims about system value should be empirically validated, and the constraints developers articulated provide testable criteria for doing so.

The use of LLMs as coders introduces the risk of identifying plausible themes that do not fully reflect participant intent. Our design addresses this in multiple ways. 
Each model proposed themes with supporting participant IDs, anchoring every candidate in specific responses. Three models from distinct providers independently discovered themes; only those corroborated across models and sufficiently evidenced in the data were retained. Two researchers then reviewed every proposed theme against its cited responses before codebook approval, removing weakly substantiated ones. Further, during systematic coding, models produced rationale before assigning codes, letting researchers trace and verify each decision. Finally, verbatim quotes throughout the paper allow readers to assess whether themes are grounded in participants’ responses.

\textbf{Internal validity.} This is a cross-sectional study capturing developers’ stated needs at a single point in time. The systems described are those developers wanted but did not have. If some such systems now exist, our findings provide criteria for evaluating how well they address developer needs. Self-selection remains a potential confound: respondents may hold stronger views about AI support than the broader population.

\textbf{External validity.} We studied Microsoft developers across global sites, diverse teams and roles, multiple domains, and varied processes. The systems and constraints we identify reflect recurring challenges that developers face in daily work, though their salience may differ in smaller organizations, open-source settings, or regulated industries. We therefore present an in-depth account of a large organizational context rather than a claim to represent all software engineers. Single case studies have advanced scientific discovery~\cite{flyvbjerg2006five} and produced foundational insights in software engineering~\cite{storey2019towards,kuper2004social}. Replication in other contexts remains necessary.

\section{22 Systems Developers Want Built}
\label{sec:systems}

We organize the 22 systems by task category. Each category contains a summary table, followed by descriptions of what developers intended each system to do, what makes it difficult to build, and the constraints they imposed on its behavior.
Those constraints recur consistently enough across respondents that we treat them as design requirements; they reflect deliberate boundaries around where developers locate professional responsibility.

\subsection{Development (N=353)}

353 of the 816 developers who completed the development block wrote substantive open-ended responses. The majority of them wanted help with the backlog of technical debt that accumulates in every mature codebase. What was striking was how consistently demand and caution coexisted: even developers who most wanted AI to take on this work attached explicit scope boundaries, review gates, and requirements that the system stop and surface its limits rather than invent past them.

\begin{table*}[h]
\caption{Systems developers want built: Development (N=353).}
\label{tab:development}
\small
\setlength{\extrarowheight}{4pt}
\begin{tabular}{>{\raggedright\arraybackslash}p{2.5cm} p{3.9cm} p{4.8cm} p{4.6cm}}
\toprule
\textbf{System} & \textbf{Problem it addresses} & \textbf{Example capability steps} & \textbf{Constraints \& guardrails} \\
\midrule
\midrule

\rowcolor{TekheletPurple!20!white}
Scoped-PR builder for tech debt removal &
  Technical debt adds up in mature codebases because careful maintenance at scale is grinding work that produces no visible feature;
  and tools that have scope overreach produce diffs nobody can review &
  Accept refactoring intent with explicit scope boundaries;
  inspect analogous implementations to learn repo naming and structure;
  produce atomic diffs with a configurable size threshold;
  halt with diagnostics when a step fails or confidence drops &
  No one-shot refactoring changes that are too large to review;
  no new classes, files, or out-of-scope restructuring without approval;
  surface missing domain knowledge rather than inventing it\\

\addlinespace

Embedded quality gate for code and tests &
Quality gates are missing during authorship because there’s no real-time error detection or test coverage feedback anchored to current diffs and codebase conventions &
  Compute the minimal impacted area during a save or stage;
  run deterministic checks first and synthesize findings into a common model;
  publish a concise quality summary before the PR is opened &
  Suggestions staged for explicit human approval, never silently applied;
  ask clarifying questions rather than guessing intent
  \\

\addlinespace
\rowcolor{TekheletPurple!20!white}
Trace-to-diff root cause workbench &
  Debugging time goes mostly to evidence collection---finding the right logs, aligning traces with the regression
  window, reconstructing runtime context from scattered systems---before a theory can even
  be tested: &
  Standardize the trigger into a failure signature and derive a minimal
  data-collection plan;
  create a failure timeline by collecting correlated signals by time/deployment version;
  generate root-cause hypotheses with causal stories and explicit
  confidence estimates;
  if a hypothesis survives verification, produce a patch and regression test &
  No code changes applied without explicit human review; reproduction must default to isolated sandboxed environments;
  uncertainty must be explicit, never hidden behind confident language;
  security-critical fixes require elevated human review;
  no patch presented to users until it compiles and passes specified tests \\

\addlinespace

Repository context graph for cross-file changes &
  Assistants lose the thread of a large codebase after a few turns and
  produce local edits that are globally wrong, duplicating or breaking contiguous systems &
  Parse source and build metadata into a persistent code graph updated as code changes;
  attach rationale from PR discussions, design decisions (ADRs), and linked bugs;
  compute an impact report for any proposed change before drafting a patch &
  All proposed diffs shown for explicit human review;
  no substantial refactoring or out-of-scope edits without interactive scoping;
  do not introduce design decisions diverging from existing conventions without approval;
  flag missing context instead of guessing \\

\bottomrule
\end{tabular}
\end{table*}

\subsubsection{\textbf{Scoped-PR builder for tech debt removal}}
The most prevalent want in this category, at 50.1\%, was help with the removal of tech debt. Mature codebases accumulate maintenance work that everyone knows needs doing(e.g., framework migrations, API renames, dependency chain management), but that is hard to do carefully at scale without deep codebase understanding. \pQuote{``Refactoring would be the biggest help. It's almost always tedious and well-defined, but it needs more than just find-replace. Almost always, it spans large file and multi-file workflows'' (P27)}.

Generating code that is contextualized was hard for available tools. Respondents wanted AI that follows the team's actual conventions, error-handling, and testing styles, then packages these changes as a sequence of diffs, each small enough for the human to review with confidence, stopping at the authorized boundary even when it identifies adjacent code worth cleaning up. Conciseness was a key requirement. \pQuote{``I want small PR's with incremental steps'' (P816)}. The worry was overreach: \pQuote{``I don't want AI to make substantial refactoring changes without interaction/consultation first...next thing I know there are new classes and new files'' (P40)}. When a change requires domain knowledge or touches core business logic, the system must recognize the boundary and stop: \pQuote{``If the code change requires specialized domain or tribal knowledge not found within the repository, I'd rather handle it myself'' (P211)}.

\subsubsection{\textbf{Embedded quality gate for code and tests}}

27.8\% of respondents wanted a quality gate embedded during authorship time: pre-commit bug-spotting, standards enforcement, vulnerability detection, and missing test identification. However, available tools provided feedback that was generic, shallow, and often unanchored to the current diff: \pQuote{``Stop giving bad advice. So many of the PR automated comments are just plain wrong'' (P609)}. Generic linter-like recommendations erode trust; what developers need is a quality pass anchored to the current diff, combining deterministic analyzers with targeted model reasoning to surface only what is genuinely worth attention. Done well, this shifts defect detection upstream: \pQuote{``you can avoid many on-call fixes if you leverage AI upstream and catch bugs during development or early test environment, before they affect any users'' (P120)}.

The constraints were strict that the systems should not bypass human accountability: no automatic modifications, no auto-submitted or auto-approved artifacts. For test generation especially, respondents wanted the system to ask before writing: \pQuote{``AI should ask for a lot of input before writing unit tests...especially if we are writing a new feature from scratch'' (P680)}. 

\subsubsection{\textbf{Trace-to-diff root cause workbench}}
19.5\% described wanting help ``assembling the case file'' for live production incidents. The bottleneck is evidence collection: an engineer has to find the right logs, identify which recent code change introduced the failure, and reconstruct enough runtime context to have something worth verifying before any debugging begins: \pQuote{``\ldots there are lots of logistics that need to be done...if those are [done] by AI, it would be a great win'' (P156)}.

Respondents described wanting a workbench that starts from a bug report, stack trace, or incident alert and assembles a debug case file: correlated logs and traces, the most likely regression window, similar historical failures, and competing root-cause hypotheses. If a hypothesis survives verification, the system needs to propose a narrow patch and a regression test. The goal of these tools is to shift engineer effort from evidence gathering to expert judgment: \pQuote{``focus on `do we agree with the Agent's assessment...and proposed fix?' and we could collectively churn through our bugs much faster'' (P120)}. Confidence calibration is what makes this hard. A wrong hypothesis stated confidently can send the team down the wrong path; \pQuote{``AI is very bad at understanding the context and throwing wrong answers with utmost confidence, makes us waste time instead of helping'' (P223)}. 
Developers were unanimous that no code changes must be applied without explicit human review, and proposed fixes must to be precisely scoped to the identified fault.
 
\subsubsection{\textbf{Repository context graph for cross-file changes}}
18.4\% of respondents reported that available AI assistants lost the thread of a codebase after a few turns, producing edits that were locally plausible but globally wrong. \pQuote{``We want/need to do a MUCH better job of analyzing a current code base, architecture, and structure so it can understand how/where to add/extend. Today it loves to duplicate code, and even break existing functionality'' (P208)}.

Respondents wanted a persistently maintained map of the code, tests, and historical discussions, so they could ask \textit{``where should this change go?''} and \textit{``what breaks if I touch this?''}  before any edits were made.
To do this, the system has to infer conventions from the existing codebase and attach rationale from PR discussions, ADRs, and linked bugs, recovering context that never made it into a commit message. Participants noted that this was harder than retrieval-augmented generation (RAG): structural understanding of the codebase and historical reasoning about its design are both required. Impact analyses have to be specific enough to act on without being so exhaustive that they are ignored. 

Scope constraints followed directly: no modifications outside the approved task, no substantial structural changes without interactive scoping, no design decisions that diverged from existing conventions without approval, and when evidence is insufficient, the system must surface the gap and ask rather than invent context.

\begin{takeawayBox}
\textbf{Takeaway}. The shared need across all four systems is context representation and appropriate scoping: each requires the AI to assemble and maintain local, history-laden tacit knowledge (e.g., repository conventions, diff intent, failure evidence) that available tools either lose between turns or never acquire.
\end{takeawayBox}

\subsection{Design and Planning (N=223)}

Of 548 developers who completed this block, 223 wrote substantive open-ended responses. They wanted AI to assist with the exploratory overhead surrounding design work while keeping the actual decisions firmly with them: \pQuote{``AI's system design solutions bias toward old known solutions rather than a modern solution that solves the problem better'' (P195)}.

\begin{table*}[h]
\caption{Systems developers want built: Design \& Planning (N=223).}
\label{tab:design}
\small
\setlength{\extrarowheight}{4pt}
\begin{tabular}{>{\raggedright\arraybackslash}p{2.3cm} p{4cm} p{4.9cm} p{4.6cm}}
\toprule
\textbf{System} & \textbf{Problem it addresses} & \textbf{Example capability steps} & \textbf{Constraints \& guardrails} \\
\midrule
\midrule

\rowcolor{TekheletPurple!20!white}
Design-to-sprint workbench &
  Turning an accepted design plan into a structured backlog, dependency map, and sprint sequence is repetitive work that consumes time without requiring much human judgment &
  Start from the user-selected scope, marking assumptions and out-of-scope items explicitly;
  generate a work breakdown tied to the team's planning template;
  infer dependencies and estimate ranges from comparable historical tasks;
  draft milestone sequencing from estimates/capacity, surfacing trade-offs &
  All changes require explicit human approval;
  must not autonomously replan or direct the team's process;
  must not assign work to individuals automatically;
  must not contact colleagues or send messages without authorization;
  must not invent missing requirements---distinguish guesses from grounded outputs \\

\addlinespace

Architecture studio for requirements-to-design &
  Design sessions often produce alternatives without a structured comparison; teams lack a tool that maintains design state (goals, constraints, exploration space), generates alternatives grounded in local context, and tracks rationale for the choices &
  Create a design state by retrieving local context from ADRs and API/infrastructure catalogs; 
  generate distinct architecture candidates with explicit trade-off comparisons;&
  Must not make or settle the final architectural decision;
  must not silently invent requirements, assumptions, or generic patterns;
  must not produce a one-shot design without gathering context and allow user steering \\

\addlinespace
\rowcolor{TekheletPurple!20!white}
Design analyzer for trade-off and risk &
  Design reviews surface obvious issues but miss hidden dependencies, Non Functional Requirements gaps, and failure modes that require systematic cross-checking &
  Organize the review around system components, data flows, trust boundaries, and open question;
  map the design to requirements, marking coverage gaps;
  cross-check against infrastructure catalogs to expose hidden dependencies;
  generate what-if scenarios and trade-offs with provenance &
  Must not present or settle the final design decision;
  must provide transparent reasoning and provenance for every critique and inflection points, never unsupported assessments \\

\addlinespace

Decision context and provenance graph &
Design rationale and its provenance live in the heads of engineers who may have moved on; decisions made months ago cannot be traced, and their downstream consequences are invisible &
Standardize artifact corpus as time-stamped records with stable links and freshness metadata;
  extract decisions, alternatives, owners, and rationale from the artifact corpus;
  link parent and child decisions so upstream choices can be traced to their consequences;
 &
  Must not make final decisions or own the plan;
  must not fabricate context---distinguish evidence from guesses and surface temporal provenance;
  every answer must expose its source; respect sensitive data boundaries \\

\addlinespace
\rowcolor{TekheletPurple!20!white}
Full-loop design doc and diagram workbench &
  Engineers, even with clear technical ideas, struggle to produce consumable documentation; the gap between understanding something and writing it up well consumes time without adding insight &
  Let the author select which inputs shape the draft and map them to required sections, label each as supported, inferred, or awaiting confirmation; generate diagrams in both rendered and structured text form to afford in-situ editing;
  support targeted regeneration of sections without rewriting the whole &
  Must not make final design decisions or silently assume missing requirements;
  must not produce a one-shot unsteerable document;
  must expose source context for every suggestion so users can verify and control output \\

\bottomrule
\end{tabular}
\end{table*}

\subsubsection{\textbf{Design-To-sprint workbench}}
30.5\% of respondents wanted a full-cycle AI workbench that absorbs the planning and coordination overhead that follows once a design is settled. For example, P380 said: \pQuote{``It would be helpful if AI could take high-level features and break them down into discrete implementation tasks, and then plan those tasks out'' (P380)}.  The system they wanted would take an approved plan and produce a first-pass execution package---hierarchical work items, dependency mapping, estimates inferred from comparable historical tasks, and a draft sprint sequence for human review---then, once the team edits and accepts them, it would sync approved deltas into the tracker and synthesize status digests from tracker activity, CI signals, and open blockers.

The constraint respondents drew most consistently was that the system should not own the plan: \pQuote{``I want to use AI to help distill and gather information\ldots\ the end decision and plan should be my responsibility'' (P11)}. It should not create or modify work items without preview and explicit human approval, assign work to individual developers, contact colleagues, send status messages without explicit authorization, or invent missing requirements.

\subsubsection{\textcolor{black}{\textbf{Architecture studio for requirements-to-design}}}
\pQuote{``I want AI to assist with exploring design alternatives, identifying edge cases, and helping align architecture with evolving business goals. It should act as a thoughtful collaborator, not just a generator'' (P249)}. 27.8\% of participants wanted such an \emph{interactive design partner} that maintains explicit design state---goals, open questions, history of exploration space---while generating multiple, materially different architectures grounded in local context (e.g., ADRs, infra- catalogs), asking clarifying questions when missing information affects the design, and tracking why each option was introduced or discarded.

The hardest part is grounded context. Participants observed that systems relying on generic patterns were of limited usefulness: \pQuote{``AI would probably fall back to using the same known pattern regardless of the design particularities. That makes the designs inefficient” (P359)}. More concerning were hallucinations, which undermine trust: \pQuote{``I don't trust the AI to design anything; it hallucinates” (P456)}. 

Participants emphasized retaining agency in making architectural decisions. \pQuote{``AI should not settle the final decision. Because this is the part that people should be accountable for'' (P774)}. Additionally, they did not want systems to produce one-shot designs without sufficient context and guidance.

\subsubsection{\textbf{Design analyzer for trade-off and risk}} 20.6\% wanted AI to analyze a proposed design the way a careful reviewer would: \pQuote{``If it can understand complex system architectures, and provide insights into gaps in your design, that would be extremely helpful'' (P83)}. They wanted AI to reconstruct component models from docs and diagrams, check requirements and NFR coverage, surface hidden dependencies, run what-if scenarios, and turn the result into a trade-off matrix, pointing back to evidence so the developer could judge whether it was correct. P2 emphasized, \pQuote{``I’d like AI to reason over the choices that I made, the impact they had, and how small changes to initial design decisions could have a significant impact on the architecture—be it complexity, performance, or cost” (P2)}.

Participants wanted to retain decision control: \pQuote{``I don't want AI to own critical decisions or trade-offs, as these require deep domain knowledge, long-term vision, and accountability that only experienced engineers can provide” (P249)}. When context was incomplete, they preferred the system to identify the decision point, explain the uncertainty, and request clarification rather than generate unsupported assessments.

\subsubsection{\textbf{Decision context and provenance graph}}
19.7\% wanted a system that could recover the context behind past decisions: \pQuote{``there can be a substantial time gap between when a stakeholder mentions an important detail and the documentation is produced and reviewed, and with the volume of topics discussed and documented, it's very easy for crucial design decisions and context to get lost or overlooked'' (P40)}. They expected the system to ingest code history, docs, tickets, notes, and opt-in conversations as time-stamped facts and decisions, and answer questions like “why was this done?”, “what changed after the RPO target moved?”, and “which later components depend on that decision?” by presenting a traceable decision lineage linked to original artifacts.

This was seen as more challenging than simple text retrieval. The system must reason over time, track provenance, indicate when sources may be stale, and explicitly surface gaps: \pQuote{``I don't want AI to give me answers without context. If I can't tell why it's suggesting [something], then I also can't tell if it's hallucinating or not” (P483)}. Participants wanted judgment to remain theirs.

\subsubsection{\textbf{Full-loop design doc and diagram workspace}}
15.2\%  wanted a workspace to transform rough materials into consumable documentation. \pQuote{``Essentially, supplement my ability to get my ideas documented in a consumable way for others. This would really help and reduce a lot of time that feels wasted to me” (P483)}. The system would take notes, transcripts, rough outlines, and selected code diffs and turn them into structured design documents and editable diagrams, each carrying provenance so the author can trace its source and refine what needs work. 

Participants noted that first-draft generation worked adequately for well-specified inputs. \pQuote{``Templatization and summary from notes and meeting context into a skeleton design with high accuracy would be a major step forward” (P387)}. But the harder problem was iterative refinement with user steering---making targeted updates to specific sections without regenerating the entire document, updating diagrams without overwriting unchanged parts, and maintaining alignment between text and visuals as they both evolve. They were cautious about systems that assumed missing requirements or introduced unsupported choices, and instead wanted the source context exposed for verification and control.

\begin{takeawayBox}
    \textbf{Takeaway.} 
The shared need is for AI systems to maintain grounded, stateful reasoning over an evolving design context. Each system required maintaining explicit design state, tracing decisions over time, and supporting iterative refinement, while keeping outputs human-gated.
\end{takeawayBox}


\subsection{Quality and Risk Management (N=155)}

155 of the 401 developers who took this block wrote substantive open-ended responses. They described a consistent gap between what automated tooling currently checks and what it would need to check to be useful: coverage at the point of authorship rather than post-merge, findings tied to the specific diff rather than the whole codebase, and risk signals calibrated to a codebase's failure history rather than generic thresholds. Equally consistent was where they drew the line: no system in this category was permitted to make a judgment, approve an artifact, or attest to a result.

\begin{table*}[h]
\caption{Systems developers want built: Quality \& Risk (N=155).}
\label{tab:quality}
\small
\setlength{\extrarowheight}{4pt}

\begin{tabular}{>{\raggedright\arraybackslash}p{2.6cm} p{3.7cm} p{5.4cm} p{4.4cm}}
\toprule
\textbf{System} & \textbf{Problem it addresses} & \textbf{Example capability steps} & \textbf{Constraints \& guardrails} \\
\midrule
\midrule

\rowcolor{TekheletPurple!20!white}
Change-aware test generation and quality gates &
  Test suites lag behind code, leaving coverage gaps as behavior changes outpace test authoring &
  Ingest diffs to build an impact map of changed artifacts;
  recover expected behavior from requirements and tests, requesting input when intent cannot be inferred;
  generate repository-style tests using existing helpers;
  publish a quality report highlighting untested changes &
Must not self-certify the correctness or completeness of its own tests and output;
  must not automatically commit generated tests or code;
  must provide auditable evidence for all activities;
  must ask and not infer missing context. \\
 
\addlinespace
Context-aware pull request review assistant &
Large pull requests (PRs) are hard to review: intent is unclear, risk boundaries are implicit, and mechanical churn obscures meaningful changes. Generic automated comments add noise while missing repo-specific concerns. &
Ingest PRs to build a layered summary of intent, affected modules and dependencies, and their downstream impact;
analyze changes for performance and maintainability using repo norms and historical review patterns;
attach findings to exact files or lines with severity and rationale; 
publish inline annotations and a summary that lists checked categories and analysis boundaries &
Must not replace human review or act as the final authority or pass/fail gate for PRs; must not automatically change code without explicit human review;
must make uncertainty explicit and abort rather than inferring intent \\

\addlinespace

 \rowcolor{TekheletPurple!20!white}
Pre-merge security advisor with patch suggestions &
Current tools flag issues, but vulnerability findings arrive late, lack context and actionable fixes, requiring significant effort in validation of these issues and their remediation. 
&
Triage at the diff level with remediation options tied to the relevant security-sensitive surfaces in the changed code; run static, taint-style, and dependency advisory checks correlated into findings tied to exact paths; generate remediation options with a conservative patch;
  stop and surface uncertainty when confidence is low &
  Must never auto-merge or auto-commit security fixes;
  must not become the sole authority for security decisions;
  must provide verifiable evidence of what was analyzed;
  must not expose sensitive information in outputs;
  must surface uncertainty when confidence is low \\
 
\addlinespace
 
Compliance evidence compiler \& interpreter &
Compliance work is dominated by evidence chase and translation: decode policy language, determine applicability, hunt down artifacts across supply-chain scanners and repos, then restate findings in auditor-expected formats. 
&
  Parse the selected policy set into required evidence types and applicability rules; collect a minimal set of scoping facts that determine which policies apply, and record the reasoning for each applicable or non-applicable judgment; 
 harvest artifacts into an evidence ledger with provenance;
  stop and request human input when evidence is missing 
  &
  Must not handle raw customer or user data during evidence collection; must not provide compliance/remediation attestations without explicit human approval;  must not make changes to production systems \\
 
\addlinespace
 \rowcolor{TekheletPurple!20!white}
Change risk radar for proactive regression warning &
Regression risks hide in signals that are fragmented across systems, rollout events, and prior incidents. On their own, they look harmless; problems surface when the relevant signals are combined, which is difficult for humans to do on their own.
 &
  Assign a canonical ID to each deployment or config event so telemetry ties back to a specific rollout;
  learn service-specific baselines and detect low-severity deviations;
  correlate deviations to recent changes using timing and historical incident similarity;
  generate a short risk analysis brief with explicit drivers &
  Must not make final ship/no-ship or risk-acceptance decisions;
  must not autonomously escalate incidents or trigger crisis communication;
  must remain read-only with respect to production environments;
  risk scores must be explainable and auditable. \\
 
\bottomrule
\end{tabular}
\end{table*}

\subsubsection{\textbf{Change-aware test generation and quality gates}}44.5\% of quality and risk respondents wanted help with a problem that P64 captured plainly: \pQuote{``Testing: tedious.''} Writing meaningful tests for changed behavior lags behind shipping the changes, so test coverage gaps accumulate. The core difficulty was figuring out which assertions mattered for a change, against behavior that was rarely documented and/or test coverage that was already sparse. 

They wanted AI to derive an impact map from the diff (changed functions, touched interfaces, downstream dependents), link requirements and existing coverage, generate the missing tests, run them, and publish a compact quality gate report highlighting modified behavior that remained untested. \pQuote{``Over the next 1--3 years, I want AI to play a major role in proactively detecting regressions, analyzing test coverage gaps, and predicting high-risk areas in code changes so we can shift from reactive firefighting to preventive quality assurance'' (P59)}. Intent representation was what made this hard:  which signals encode what a diff is supposed to do, and which of those signals actually exist in a typical development workflow. Without intent, a test generation system produces tests that pass trivially or validate the wrong behavior.

Accountability constraints were strict. \pQuote{``We always need a person to be accountable for verifying: `what testing did the AI do? Do we have a report with screenshots, proving that the AI actually did what it claims?'\,'' (P120)}. The system must not decide what validation criteria should be, check-in generated tests automatically, or be the arbiter of its own output: \pQuote{``you must not have a GenAI system to evaluate itself'' (P359)}.

\subsubsection{\textbf{Context-aware pull request review assistant}}
23.2\% of respondents wanted a first-pass analyst in the PR workflow that understood the codebase well enough to provide meaningful feedback on a diff, identify which modules and interfaces were most affected, and attach analysis findings (e.g., defects, readability, security-sensitive surfaces) with rationale. 
The persistent problem participants noted was that in large pull requests, they spent most of their time reconstructing intent and tracing which interfaces are at risk before they can evaluate anything. Generic automated comments added noise and made it worse. Participants wanted a system that surfaced what aspects (e.g., security, maintainability) it had and hadn't checked: \pQuote{``mention in a comment that it has reviewed `xyz' aspects of the change and the confidence of the review so that human reviewers can look at other aspects'' (P429)}.

Participants emphasized that such systems should not approve PRs, automatically modify code, or act as a final authority: \pQuote{``I don't want AI to just act as a red-light / green-light. It should raise issues after/while humans are reviewing and still require human review'' (P92)}. And when intent is ambiguous, it should abort rather than guess: \pQuote{``If there is something that it isn't capable of doing, abort with that information, rather than making some assumptions'' (P350)}.

\subsubsection{\textbf{Pre-merge security advisor with patch suggestions}}
21.9\% of respondents wanted security triage at the diff level with remediation tied to the actual code path. \pQuote{``I wish it could ACCURATELY make assessments about security issues and functional correctness of code'' (P55)}. Existing tools already found the point-problems. The delay came after the detection: developers had to determine whether the finding was real in their code context, whether the vulnerable change was reachable, and what a safe local fix looked like in their specific repository. \pQuote{``I wish it could ACCURATELY make assessments about security issues and functional correctness of code'' (P55)}. A useful system had to tie findings to concrete artifacts---code, dependencies, configurations---and draft a plausible remediation the developer could review and accept or reject.

Accountability was the governing constraint. Developers wanted to retain responsibility for validating every finding: \pQuote{``I do not want AI to be solely responsible for quality and security; a developer needs to validate the results without leaning on it entirely'' (P42)}.

\subsubsection{\textbf{Compliance evidence compiler and interpreter}}
14.2\% described compliance work as draining without requiring much engineering expertise: \pQuote{``A security review today requires a hundred-question survey, much of which is querying and fetching data through a time-consuming process.'' (P210)}. The core activity was translation: decode policy language into developer steps, determine which policies applied to this system, collect artifacts across dependency/supply-chain scanners and repositories, and restate the findings in auditor-readable form. P476 expressed: \pQuote{``security and compliance is still a gigantic cluster [expletive], mostly because security folks think their vocab means a damn thing to us developers, and never leaving clear actionable steps to complying'' (P476)}. 

Participants wanted AI to translate a selected policy set into an applicability matrix, fetch required artifacts from engineering systems, and draft policy questionnaire answers with references, timestamps, and explicit gaps already marked. 

The constraint was firm: developers wanted a workflow engine for evidence collection only. They did not want it to declare a service compliant, auto-apply a remediation change, or handle raw customer data during evidence collection: \pQuote{``AI should not be allowed to handle user or customer's data as it could lead to a breach of confidential / privacy data'' (P292)}.

\subsubsection{\textbf{Change risk radar for proactive regression warning}}
11.6\% of respondents wanted AI to use telemetry and change history to flag high-risk changes before shipping: \pQuote{``Should be able to detect high-risk changes and de-risk them'' (P47)}. The problem was that regression risks appeared as weak signals spread across rollout events, telemetry drift, and prior incident patterns, and looked harmless individually. Participants wanted AI to score recent deployments and config changes against service-specific baselines and historical failure patterns, then issue risk briefs explaining why a change looked risky, which signals were drifting, and what impact radius was most plausible. The difficulty is in identifying which signals are actually predictive, how to keep false-positive rates low enough that teams trust the alerts, and how to present risk scores in a way that supports judgment.

Developers also wanted such systems to be explainable to be useful: \pQuote{``I do not think AI can be used to evaluate risk because humans cannot make sense of what the AI is doing 'underneath the hood' to produce its output'' (P236)}. They did not want it to make risk-acceptance decisions or escalate incidents or trigger crisis communication autonomously: \pQuote{``I don't want it making autonomous decisions about incident escalation or crisis communication without human approval'' (P740)}.

\begin{takeawayBox}
\textbf{Takeaway.} The shared need is left-shifted quality assurance: catch defects, coverage gaps, and risk at authorship time, where the cost to fix is lowest. In every case, the system detects and recommends; it does not approve, attest, or escalate.
\end{takeawayBox}

\subsection{Infrastructure and Operations (N=101)}
101 of 283 block completers answered the open-ended questions substantively. The described infra-work as toil-heavy (e.g., grunt alert triaging, pipeline maintenance, support screening). They wanted AI to absorb that toil while keeping production write access firmly behind human gates. That boundary was non-negotiable. As P476 put it, \pQuote{``AI is not a system administrator. It does not get any permissions to do anything but read and alert.''} 

\begin{table*}[t]
\caption{Systems developers want built: Infrastructure \& Ops (N=101).}
\label{tab:infra}
\small
\setlength{\extrarowheight}{4pt}
\begin{tabular}{>{\raggedright\arraybackslash}p{2.3cm} p{4cm} p{5.3cm} p{4.4cm}}
\toprule
\textbf{System} & \textbf{Problem it addresses} & \textbf{Example capability steps} & \textbf{Constraints \& guardrails} \\
\midrule
\midrule

 \rowcolor{TekheletPurple!20!white}
Telemetry correlation assistant for alert tuning and incident triage &
  Alert fatigue and incident orientation overwhelm on-call engineers; the first hour of an incident goes to assembling context that should already be assembled 
  &
  For each alert, build a compact, standardized evidence bundle from correlated telemetry, related services, and recent changes; use historical data to suggest threshold, deduplication, and coverage improvements with clear noise vs. miss trade-offs; update the dependency graph from traces to distinguish upstream causes from downstream symptoms; and produce a structured incident brief with impact, timeline, likely causes, and unknowns
  
&  
Read-only by default; no production-affecting actions without explicit human oversight; no autonomous emergency decisions, rollbacks, or security policy changes; exclude irreversible or data-loss-inducing actions from automated workflows \\

\addlinespace
 
CI/CD \& Infrastructure-as-code blueprint builder &
Pipeline definitions are hard to author and harder to debug; engineers waste hours on configuration that is poorly documented and produces opaque errors &
  Parse the repository into a delivery profile identifying build, test, packaging, and deployment targets;
  construct a topology map of CI/CD stages, jobs, gates, and dependencies;
  generate CI/CD skeletons compliant with project \& org. templates;
  produce minimal migration diffs for legacy definitions with staged rollout notes &
  Read-only against infrastructure; no auto-deploy, auto-approve, or live write permissions;
  all changes output as reviewable artifacts, not execution;
  generated artifacts must be reproducible and fully inspectable \\
 
\addlinespace
 \rowcolor{TekheletPurple!20!white}
 Maintenance backlog prioritizer &
  Deprecations, security findings, and runtime drift accumulate across services without a prioritized, actionable agenda; engineers do not know what to fix first or why it matters &
  Map services to environments, owners, runtimes, and deployed artifacts; normalize upkeep signals into a unified remediation queue; deduplicate repeated alerts into root remediation actions; prioritize by severity, impact, and effort with rationale &
  Read-only with no default administrative permissions; no autonomous patches, upgrades, or security config. changes in production
 \\
 
\addlinespace
 
Customer support triage assistant &
Repetitive tickets drain time because every case requires telemetry collection and repository searches before teams can confirm whether it’s an already existing problem or a fixed issue &
  Organize the request, redact sensitive fields, and attach relevant privacy policies before retrieval begins;
  classify the ticket by product area, issue type, and severity with rationale snippets;
  retrieve similar resolved cases and approved knowledge sources;
  draft a triage card and a customer-safe reply for human review &
  Must not send messages autonomously or position AI as the primary customer interface;
  human escalation must remain easy and immediate;
  must not expose customer data or internal/production code or have write-access to these artifacts \\
 
\bottomrule
\end{tabular}
\end{table*}

\subsubsection{\textbf{Telemetry correlation assistant for alert tuning and incident triage}}
40.6\% of respondents wanted help with the orientation phase of incident response: identifying which services were affected, which deployments or configuration changes were plausible contributors, and how the situation compared to similar past incidents. They wanted a system that assembles a compact evidence bundle---correlated logs, neighboring component signals, recent change events, similar fault history---and surfaces it as a structured brief to the engineer. Beyond triage, they also wanted the system to improve alert quality over time: learning service-specific baselines from historical data to propose threshold and deduplication changes, reducing noise without increasing missed detections. The want was to shift an engineer's effort from evidence gathering to judgment. As P83 put it, \pQuote{``Automating queries in logs, making alarms smarter, reducing manual intervention by getting to root cause faster.''}

Respondents did not want production-affecting remediation or recovery actions without explicit oversight: \pQuote{``Executing operations against production resources to try and resolve incidents without human oversight\ldots\ I would not expect an AI agent to have enough background/historical context to make the correct decision'' (P639)}. Autonomous rollbacks and security policy changes were also out of scope, as \pQuote{``these actions carr[ied] high risk and require human judgment to weigh context, trade-offs, and potential impact'' (P706)}.

\subsubsection{\textbf{CI/CD and infrastructure-as-code blueprint builder}}

33.7\% described CI/CD pipeline configuration as a source of disproportionate friction: repetitive enough to template, brittle enough that a wrong assumption could take down a deployment, and organization-specific enough that public examples often taught the wrong patterns. \pQuote{``AI could be useful in helping set up [product-name] CI/CD pipelines, since the system is notoriously poorly documented and hard to figure out. Training an AI system on it could significantly reduce developer frustration''} (P214).

Two needs emerged: (1) authoring CI/CD pipeline and infrastructure definitions from repository context while respecting organization-specific policy, and (2) diagnosing build and deployment failures by tracing through layers of configuration that few engineers fully understood. Respondents wanted a system to read the repository's build targets and workflow files, explain what the current pipeline actually does, generate a baseline configuration for new services, and diagnose failure logs with concrete diagnostic notes.

The constraint was tight around execution: no production write permissions, no automatic deployment or approval without a human intermediary. \pQuote{``it's important that a human familiar with the relevant services ensures that changes proposed by AI won't introduce issues'' (P125)}. Generated artifacts had to be fully reproducible and inspectable.

\subsubsection{\textbf{Maintenance backlog prioritizer}}
16.8\% wanted support for service maintenance overhead: \pQuote{``Handling of [service compliance] alerts with AI would be absolutely huge. That would significantly reduce the engineering demand of service ownership and increase the throughput for teams'' (P500)}. Currently, deprecation warnings, security findings, runtime drift, platform notices, and cost anomalies arrive from multiple scanners without a prioritized agenda---often as duplicated alerts, without saying what to fix, why it matters now, or who owns the missing context. Respondents wanted a system that consolidates these signals into a maintenance agenda already annotated with the context needed to act: \pQuote{``What to do / Why I'm doing it (short) / How to do it, in fine detail / Why I'm doing it (long) / Who to bother if something doesn't line up.'' (P476)}.

Respondents did not want AI to execute patches, upgrades, or configuration changes in production automatically, hold administrative permissions, or modify security configurations without human authority. Recommendations had to be reviewable with evidence; rubber-stamp approval workflows were exactly the kind of oversight erosion they were guarding against.

\subsubsection{\textbf{Customer support triage assistant}}
11.9\% described that customer support teams lost time twice: first deciding which category a ticket belonged to, then repeating the same telemetry lookups and knowledge-base searches for patterns they had already handled. \pQuote{``If an AI agent could actually look at support request text and logs and make screening, bucketing, and triage decisions based on the content, that would be super helpful'' (P92)}. \pQuote{``We got a lot of repetitive customer questions, like permission issues. I hope AI can detect patterns of our customer request triaging'' (P117)}.

They wanted a system to organize incoming tickets, match them to known issue patterns, run only pre-approved telemetry lookups, and produce a triage card plus a customer-safe response draft for the human to review and send. 

Participants emphasized that customer data, proprietary code, and sensitive information required strict privacy protection and least-privilege access. Human escalation had to remain easy and immediate: \pQuote{``As a customer, being forced to an AI is frustrating. I would prefer the AI tooling to be passive, but provide me (as the person providing customer support) prompts on suggested resolutions or ways forward'' (P2)}.

\vspace{-2mm}
\begin{takeawayBox}
\textbf{Takeaway.} The shared problem is a representation gap: the right information exists, but not in the required form. Each system bridges that gap, translating artifacts into aligned documentation, personalized ramp-up paths, audience-calibrated communications, or structured option spaces. In every case, the system drafts; the human decides.

\end{takeawayBox}


\subsection{Meta-Work: Documentation, Knowledge, Collaboration (N=157)}

157 of 532 block completers answered the open-ended questions substantively. Their primary ask was for AI to absorb the overhead surrounding engineering work to mitigate documentation drift, reduce onboarding time and communication overhead, and help with early-stage research. The distinction respondents drew consistently was between logistics, which they were willing to hand over, and judgment, learning, and voice, which they were not.

\begin{table*}[t]
\caption{Systems developers want built: Meta-Work (N=157).}
\label{tab:metawork}
\small
\setlength{\extrarowheight}{4pt}
\begin{tabular}{p{2cm} p{4.3cm} p{4.9cm} p{4.6cm}}
\toprule
\textbf{System} & \textbf{Problem it addresses} & \textbf{Example capability steps} & \textbf{Constraints \& guardrails} \\
\midrule
\midrule

 \rowcolor{TekheletPurple!20!white}
DocSync: Continuous Documentation Synchronizer &
  Documentation drifts unnoticed from the code it describes; engineers rarely update docs when they ship changes, and the gap compounds until documentation is actively misleading &
  Inventory the repository's documentation surfaces and conventions; identify docs susceptible to drift based on history; compute an impact map linking changed interfaces and behaviors to docs;
  pull facts only from authoritative sources;
  generate patches for affected sections rather than rewriting documents &
  All outputs remain drafts until explicitly approved;
  prefer minimal targeted patches over bulk generation;
  must not fabricate details or present inferred rationale as fact; leave placeholders or request input when source is absent. \\
 
\addlinespace
 
Contextualized Developer Ramp-Up Coach &
  Onboarding engineers spend weeks orienting to a new codebase with documentation that is generic, outdated, or inaccessible; the time cost falls on both the new engineer and the team members they interrupt &
  Capture the learner's goal, skill level, and constraints;
  generate an onboarding playbook with explicit human touchpoints for mentoring;
  produce a codebase map and curated reading path in the order that matters for the learner's goal;
  create hands-on exercises tied to the actual target repo and adapt based on completed work &
  Must not replace human-led mentoring or team integration---technical ramp-up is its scope; guidance must carry citations and surface uncertainty; must acknowledge missing context rather than inferring; generated materials require human review \\
 
\addlinespace
 \rowcolor{TekheletPurple!20!white}
Stakeholder Communication Draft Workbench &
  Tailoring technical updates for different audiences is repetitive work; engineers with clear information still spend significant time on phrasing, audience adaptation, and register. &
  Start from a drafting action where the user specifies the audience, intent, technical fluency, language, and formality expectations;
  extract facts, deltas, open risks, and requirements from the chosen artifacts;
  generate drafts at the right technical depth without flattening important nuance &
  Must never send messages or share generated communications automatically;
  must not handle sensitive or relationship-critical communications;
  must enforce strong privacy boundaries, especially when external recipients are involved\\
 
\addlinespace
 Interactive Exploration Board For Tech Discovery &
 
Before architecture begins, there is a fuzzier phase that does not yet look like design: someone has a problem, an instinct, and maybe one favored approach. Current tools arrive too late, after the option space has already narrowed, and hand back polished answers that reinforce the first instinct rather than stress-testing it. The result is commitment before exploration &
 
Start from a research question and capture constraints, assumptions, and evaluation criteria; generate multiple option cards with context and explicit assumptions; score competing approaches against user-chosen criteria; let the user reweight criteria, add candidates, or challenge the current set; spin up disposable prototype spikes or diagrams for options worth testing
&
Must not make design decisions or inject unsolicited ideas into an active thinking process; must not treat the first option set as exhaustive; all claims must carry explicit uncertainty markers and require human validation before acting on them.\\
  
\bottomrule
\end{tabular}
\end{table*}

\subsubsection{\textbf{DocSync: continuous documentation synchronizer}}
45.9\% of respondents wanted help keeping documentation from drifting from the code it described. \pQuote{``I think the biggest role that AI can assist in is maintaining documentation to be accurate and up-to-date. Missing, incomplete, or outdated documentation is arguably the biggest pitfall of development" (P129)}.

They wanted systems that monitor code changes, detect which documentation is affected, and propose precise updates grounded in the modified code, tests, and schemas. As one noted, \pQuote{``Unfortunately, I think right now the focus is on just generating a bunch of AI slop that nobody will ever read or maintain, which just makes all the documentation problems we have now even worse'' (P456)}.

Developers emphasized that all outputs must remain drafts pending explicit human approval. These systems should not fabricate details when evidence is missing or present inferred rationale as fact; they should leave placeholders or request input: \pQuote{``Auto-documentation of 'what' would be great. AI models aren't smart enough to completely explain the 'why's yet'' (P171)}.

\subsubsection{\textbf{Contextualized developer ramp-up coach}}

30.6\% wanted help accelerating the onboarding process for new engineers. P380 described: \pQuote{``It would be helpful if AI could take a new engineer's goals and skill level and generate a structured onboarding path, rather than leaving them to piece it together from outdated docs'' (P380)}.
They wanted a system that captures the learner's goal, skill level, and time constraints, then composes a role-specific ramp-up path from the actual repositories, docs, ADRs, setup scripts, and examples the learner would use. It would generate a codebase map, curated reading path, and hands-on exercises with source-linked explanations, adapting as work was completed and flagging explicit human touchpoints for mentoring and team integration. As one participant noted, \pQuote{``AI can help a lot in onboarding new people into a team, and help in speeding up the time for them to start contributing to the project” (P781)}.

The constraint respondents drew consistently was that the system must not replace human mentorship or team integration: \pQuote{``Onboarding requires human involvement to feel welcomed and a part of the team" (P281)}. AI could structure the logistics of onboarding, but culture and relationship-building remain with the humans.

\subsubsection{\textbf{Stakeholder communication drafting workbench}}
12.7\% wanted help with the overhead of tailoring technical updates for different audiences. \pQuote{``I would like AI to help with tailoring stakeholder communications to different stakeholders. Right now, so much team meta-effort goes into preparing tailored communications'' (P40)}. 
Participants wanted AI-assisted drafting to begin with explicit audience configuration, including intent, technical fluency, language, and formality expectations, after which the system would extract facts, deltas, open risks, and requirements from selected artifacts and generate drafts calibrated to that context, preserving technical nuance. For multilingual communication, participants wanted targeted language coaching that preserved their own intent.

Accountability was the most frequently cited boundary: \pQuote{``AI should not write communications to the customer, or anyone else---its writing style is so obvious, and it makes us look lazy'' (P457)}. No message should be sent automatically; every output must remain a draft until approved, and sensitive or relationship-critical communications (those requiring empathy, trust-building, or nuanced trade-off discussion) must stay with the human.

\subsubsection{\textbf{Interactive exploration board for tech discovery}}

11.7\% described a gap in the earliest phase of technical decision-making---the fuzzier research phase---where someone has a problem, an instinct, and maybe one favored approach, but no structured way to research the trade-offs between alternatives. Available tools arrive too late, after the option space has already narrowed, and hand back polished answers that reinforce the first instinct rather than stress-testing it. As P457 put it, \pQuote{``Too often we just have one idea---hey, I want people's opinions, I'll send out a meeting request, but that ends up being a big ordeal.''}

11.7\% wanted a system that operated in this earlier phase: starting from a research question with explicit constraints and evaluation criteria, generating materially different option cards with local precedent and explicit assumptions attached, and letting the engineer re-weight criteria, add candidates, or challenge the current set before any commitment was made.

The constraint was preserving the integrity of early thinking. The system has to be pull-based: never injecting unsolicited options into an active exploration, and making explicit that its outputs were a starting point, not a conclusion. As P18 put it, \pQuote{``AI can be OK as a sounding board for research, as long as there is human validation of what it says and the user remembers that the AI can be wrong.''}

\begin{takeawayBox}
\textbf{Takeaway.} The target of these four systems is the orientation cost before work can begin, assembling signals that are spread across too many sources. Each system addresses a different signal type, but the shared design is the same: collate, prioritize, and surface. Final accountability stays with the human.
\end{takeawayBox}

\section{Discussion}

\subsection{Implications for practice}

\subsubsection{Bounded Delegation: Protecting the Parts Worth Doing.}
Reflect on your own daily work for a moment: \textit{Which parts would you want an AI to do for you? Which would you insist on doing yourself, even if the AI could do them independently?} 

We call this split \textit{``bounded delegation''}: developers in our study expected AI to support work that was tedious, time-consuming, and context-heavy, but not to subsume their \textit{craft} or \textit{judgment}. There are two parts to this boundary, and only one is likely to move.

The first reflects current tool limitations. Developers described existing tools as unreliable enough to warrant caution in delegation: \pQuote{``AI is very bad at understanding the context and throwing wrong answers with utmost confidence'' (P223)}. As models improve, this part may shift.

The second reflects agency. \pQuote{``AI should not settle the final decision ever. Because this is the part that people should be accountable for'' (P774)}.
The architecture studio is meant to generate alternatives but not decide; compliance tools to surface gaps but not sign off; code review assistants to identify issues but not approve changes. These are intentional boundaries on what developers choose to retain, regardless of model capability.

Hackman and Oldham's work-design theory offers an explanation~\cite{hackman1976motivation}: task identity and agency are among the core drivers of meaningful work. When AI absorbs craft, it risks eroding a developer's sense of owning a coherent piece of work. Bounded delegation may thus be a developer's way of preserving that sense, shaping AI's role to protect what makes the work worth doing.

\begin{implicationBox}
    \textbf{Key implication}: AI systems should be designed around bounded delegation: maximize support for assembly while preserving human craft and agency. Treating this boundary as a design requirement, i.e., an enduring feature of how humans and AI systems share work, enables complementarity~\cite{shneiderman2020human} that sustains both productivity and the satisfaction that makes it durable.
\end{implicationBox}


\subsubsection{The Right-Shift Problem}

When a service goes down at 3 AM, the on-call engineer has to fix it. Historically, you could wake up the original author, check the commit history, or rely on team knowledge to understand the logic. Today, engineers are increasingly asked to fix code that no one on the team fully understands. 

This reflects a right-shift problem in the software lifecycle. As code generation becomes easier, verification can become the constraint. Reviewers often need to reconstruct intent, assess correctness, and identify risks across larger volumes of machine-generated code with limited provenance or design context.

The consequences can be cumulative. Faster generation does not necessarily lead to better systems; instead, effort shifts downstream~\cite{afroz2025developer}. The resulting flow of AI-generated output is already outpacing existing verification practices~\cite{pearce2025asleep}, increasing the cognitive cost of review and, increasingly, contributing to developer fatigue and burnout~\cite{miller2025maybe, feng2025gains}. Over time, this imbalance can also lead to the accumulation of AI-induced technical debt~\cite{liu2026debt, choudhuri2025needs}, making systems harder to reason about, maintain, and trust.

The systems identified in this study respond to this imbalance by moving verification earlier in the workflow. Developers consistently asked for systems that surface quality signals at the point of change: be it change-aware testing that highlights coverage gaps at the diff, or pre-merge security analysis that identifies risks before deployment. Each targets the same moment: exposing context and intent during authorship, while the developer still has enough proximity to act.

This emphasis on left-shifting can also be seen as a work-design intervention. Embedding earlier, more targeted feedback loops into daily work can keep developers closer to the consequences of their decisions, preserving quality judgment over time.

\begin{implicationBox}
    \textbf{Key implication:} AI systems should be designed to scale verification commensurate with generation by left-shifting quality signal provision. Without corresponding support, impetus on faster production causes sustained pressure on developers and makes systems harder to understand, maintain, and trust.
\end{implicationBox}

\subsubsection{Cross-Cutting Guardrails}

All 22 systems came with constraints under which developers would accept it. Across all categories, four constraints recurred often enough that we treat them as guardrails for acceptable system behavior: (1) explicit authority scoping: operate within a declared boundary and halt when judgment is required; (2) provenance: trace outputs to sources; (3) uncertainty signaling: surface when evidence is missing or confidence is low; and (4) least-privilege access with strict boundaries around sensitive data.

These guardrails reflect a view of how agency should be structured, regardless of what any AI system can do. As developers offload assembly work to AI, they require that responsibility remain visible, attributable, and interruptible. In this sense, \textit{guardrails operationalize bounded delegation}: they define how work can be delegated without relinquishing ownership of outcomes. This also bears directly on the \textit{right-shift problem}. As AI produces more outputs, provenance and uncertainty signaling are what let humans keep up with verifying them.

What remains an open question is calibration. Too little constraint and developers lose meaningful control and trust. Too much, and they route around the systems entirely. Where this boundary lies likely varies by context and task criticality. We see getting this right as a priority for future work.

\begin{implicationBox}

\textbf{Key implication:} 
Authority scoping, provenance, uncertainty signaling, and least-privilege access should be treated as design requirements. The challenge is calibration: enough to preserve agency, not so much that developers bypass the tools.

\end{implicationBox}

\vspace{-3mm}

\subsection{Implications for research}
\label{sec:research}
Using generative AI models in a qualitative analysis pipeline sits in tension with traditional reflexive thematic analysis, which depends on human interpretation, theoretical sensitivity, and judgment~\cite{braun2006using,braun2022conceptual}. Our pipeline does not claim that these models can replace interpretive qualitative work. Instead, it separates two activities that are often intertwined in large-scale analysis: (1) \textit{pattern discovery across a large corpus} and (2) \textit{interpretive judgment about what those patterns mean}. The models supported the first; the researchers retained authority over the second.

This distinction was enforced throughout the pipeline. Models proposed candidate themes with supporting participant IDs, but every theme in the final codebook passed through a human review gate in which two researchers examined the cited responses, assessed fit, clarified boundaries, split or merged themes, and added concepts the models missed. During systematic coding, models were restricted to the approved codebook and required to provide rationales before assigning codes, making their decisions reviewable rather than opaque. 
In this sense, the models served as discovery and coding instruments under human supervision; interpretive authority was never delegated.


From this experience, we draw three practical design principles for researchers considering similar workflows. First, \textit{separate discovery from interpretation}: allow models to discover candidate themes from the data, but finalize the codebook under human supervision before systematic coding begins. Second, \textit{require evidence and rationale before codes}: models should cite supporting responses during theme discovery and provide rationale during coding so that researchers can inspect whether outputs are actually grounded in the data. 
Third, \textit{use model diversity as a safeguard}: models from different provider families surface different blind spots and failure modes, making agreement more meaningful and disagreement more diagnostically useful.

These practices offer a practical path for using generative AI models in large-scale qualitative studies without collapsing responsibility for interpretation or rigor. They, however, do not resolve the deeper epistemological question of whether machines can interpret human meaning~\cite{schroeder2025large}, nor do we sidestep it. 
What we argue is that making the boundary between AI assistance and human interpretation explicit and auditable at every stage in qualitative analysis is, itself, a methodological contribution.

\begin{implicationBox}
    \textbf{Key implication}: Separate theme discovery from coding, keep interpretive authority human-gated at every stage, require auditable rationale throughout, and use model diversity as a convergent validity check when using generative AI in qualitative analysis pipelines.
\end{implicationBox}

\section{Conclusion}

We surveyed 860 Microsoft developers to identify what AI systems they want built for their daily work. The result is a catalog of 22 systems across five task categories, each described in terms of the problem it solves, what makes it hard to build, and the constraints developers place on its behavior.
These systems sit overwhelmingly on the verification side of the workflow, reflecting the current bottlenecks in AI-assisted development. Across these systems, developers enforced four recurring constraints: explicit authority scoping, provenance, uncertainty signaling, and least-privilege access.

Our findings reveal an underlying pattern we call \textit{bounded delegation}: developers drew a consistent line between the overhead they wanted AI to absorb and the craft they refused to let go. We argue this should be treated as a durable design requirement, one that reflects where developers locate ownership and meaning in their work.
\textit{How, then, should we define the value of AI in software engineering: by how much work it can do, or by where, and how precisely, it stops?}

\vspace{2mm}

\textbf{Data availability.} The supplemental package, including method artifacts, codebooks, and system cards, is available at~\cite{supplemental}. The project landing page is at \url{https://aka.ms/AI-Where-It-Matters}.

\bibliographystyle{ACM-Reference-Format}
\bibliography{bibfile}

\end{document}